\documentclass[twocolumn]{revtex4-1}
\usepackage{ifthen}
\usepackage{color}
\usepackage{graphics,graphicx,epsfig}
\usepackage{epsf,epstopdf,wrapfig}
\usepackage{amssymb,amsfonts,amsmath}
\usepackage[export]{adjustbox}
\include{epsf}
\usepackage{ifthen}

\newcommand{\beqn}{\begin{eqnarray}}
\newcommand{\eeqn}{\end{eqnarray}}
\newcommand{\beq}{\begin{equation}}
\newcommand{\eeq}{\end{equation}}

\definecolor{junglegreen}{rgb}{0.16, 0.67, 0.53}
\definecolor{myrtle}{rgb}{0.13, 0.26, 0.12}
\definecolor{lincolngreen}{rgb}{0.11, 0.35, 0.02}
\definecolor{forestgreen}{rgb}{0.13, 0.55, 0.13}

\newcommand{\rev}{\color{black}}

\newcommand{\figureone}{
\begin{figure*}
\begin{center}
\includegraphics[width=.75\linewidth]{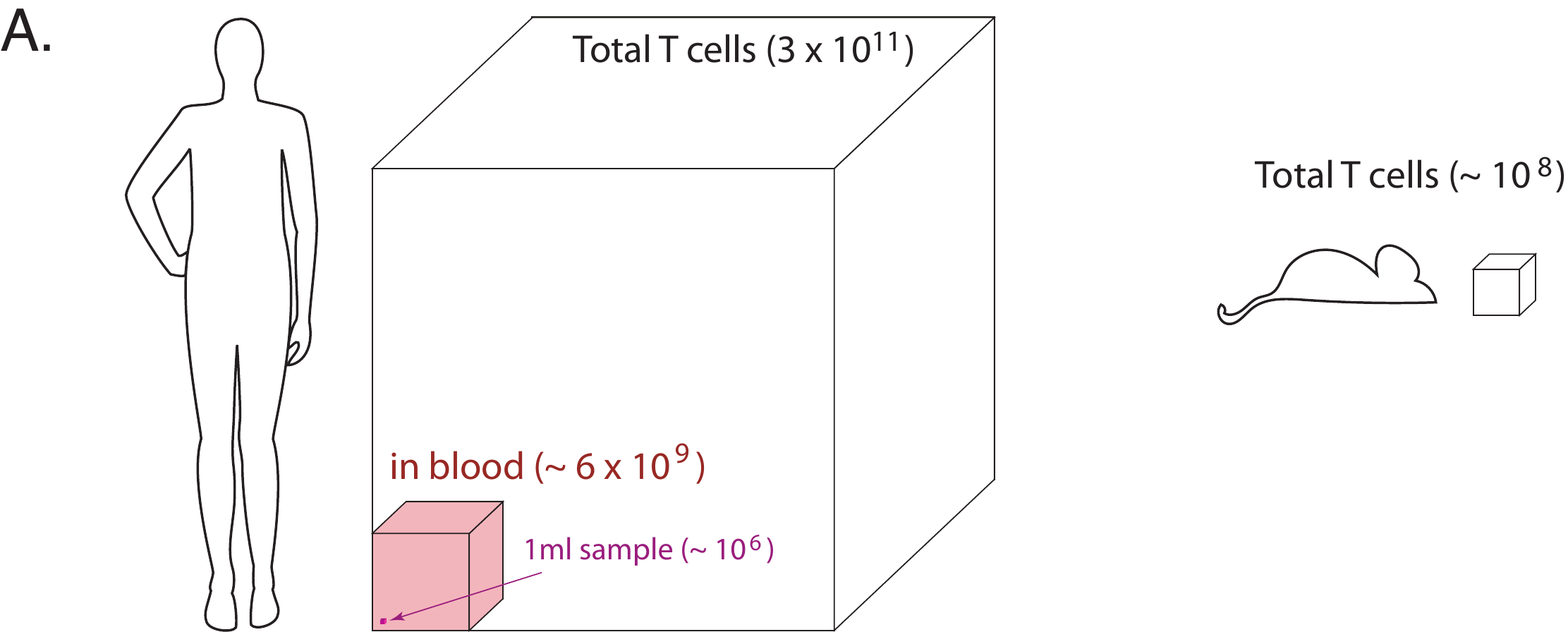}
\includegraphics[width=\linewidth]{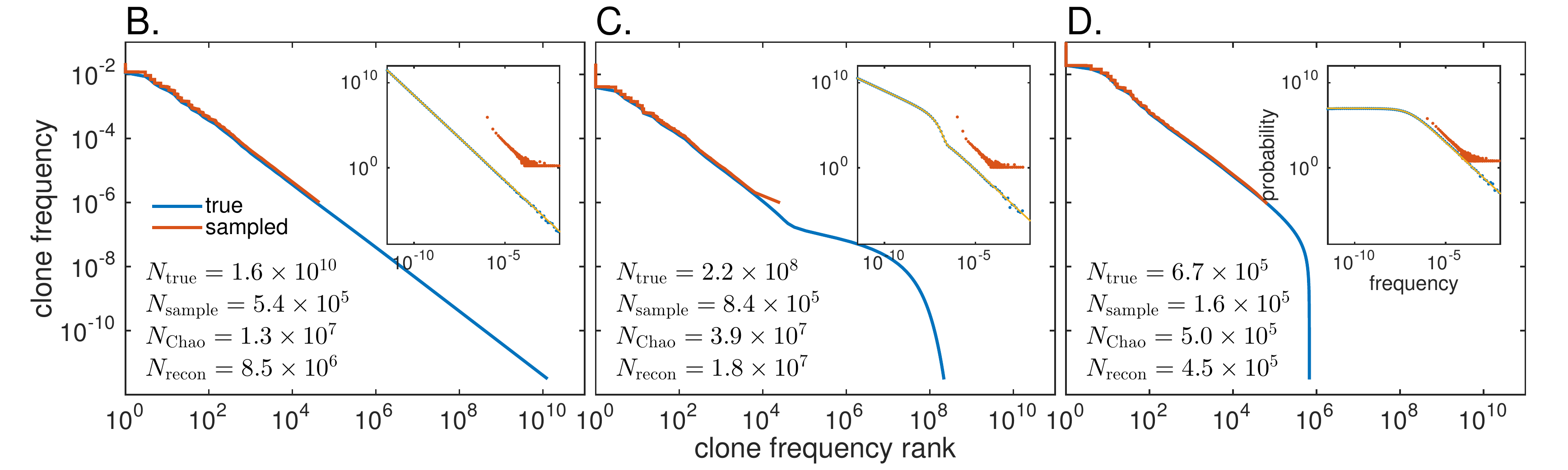}
\caption{\textbf{Estimating the total number of B or T cell clonotypes from small samples is generally impossible}. {\bf A.} Orders of magnitude for the number of T cells. Only one or a few percent of all $T=\sim 3\cdot 10^{11}$ T cells circulate in blood at any given time. Among these, typical sequenced samples contain about a million cells, which is a tiny fraction of the total repertoire. The number of T cells in a mouse is shown for comparison. Similar numbers hold for B cells. {\bf B.-D.} Rank-frequency plots of three synthetic repertoires showing the frequency of B- or T-cell clones {\em versus} their rank (from most frequent to least frequent). The corresponding clone size distribution is shown in the inset: {\bf B)} power-law distribution; {\bf C)} mixture of power-distribution and neutral distribution (see main text); {\bf D)} power-law distribution with low-frequency cutoff. A random of sample of $10^6$ cells (in red) fails to capture most of the true rank-frequency relation (in blue). While the sampled distribution looks similar in all three cases, the true distributions are very different in the domain of low clonal frequencies, and correspond to a widely different number of clones $N_{\rm true}$. That number is very poorly estimated by the number of sampled clones ($N_{\rm sample}$), or {\rev by statistical estimators such as Chao1 ($N_{\rm Chao}$) \cite{Chao1984}, or Recon ($N_{\rm recon}$) \cite{Kaplinsky2016}.}
}
\label{fig:intro}
\end{center}
\end{figure*}
}

\newcommand{\figuretwo}{
\begin{figure*}
\begin{center}
\includegraphics[width=\linewidth]{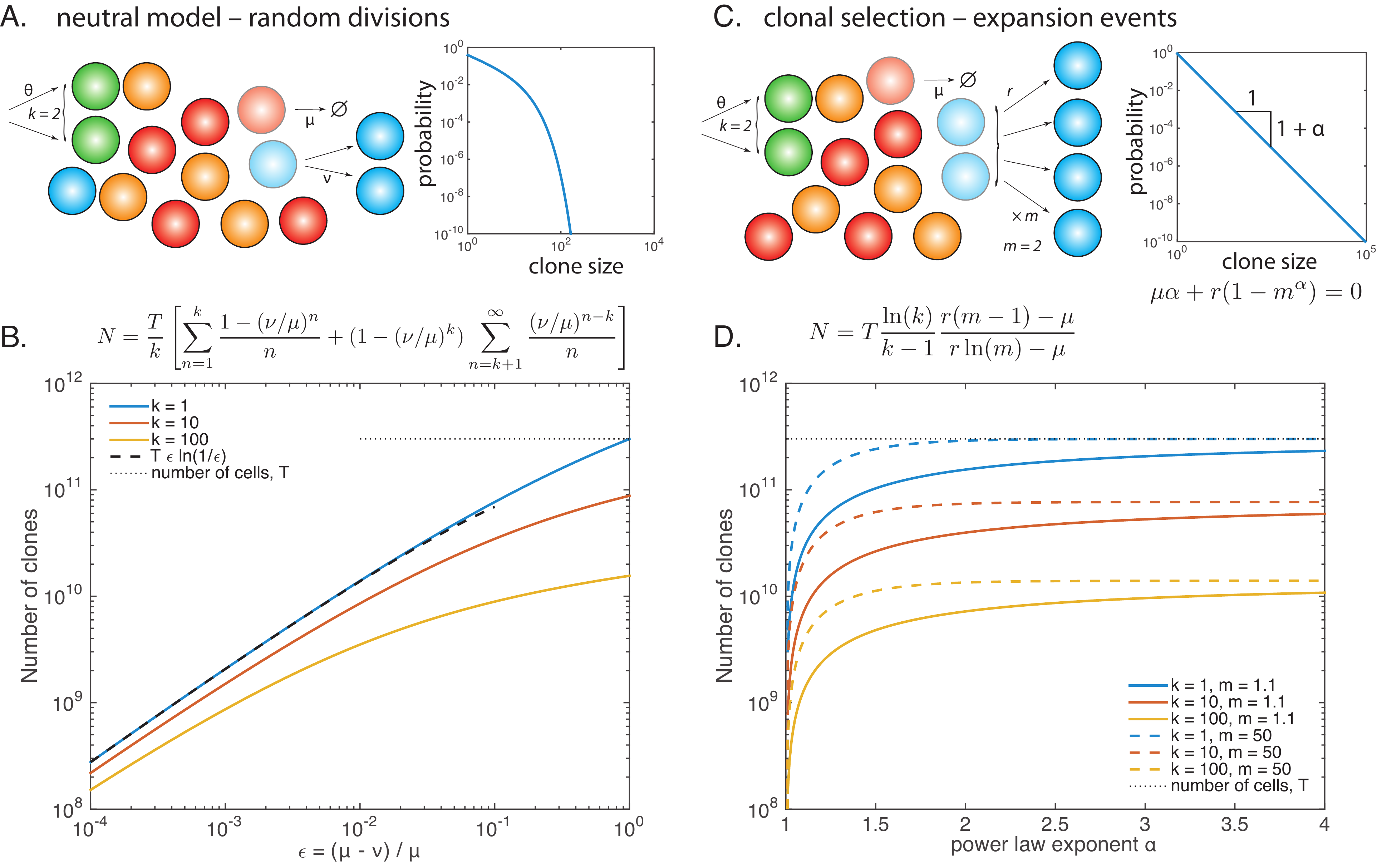}
\caption{\textbf{Lymphocyte population dynamics models can be used to estimate the number of clones. A.} Neutral model for lymphocyte dynamics. New clones come out of the thymus (for T cells) or bone marrow (for B cells) with rate $\theta$, with initial clone size $k$. Then each cell may divide with rate $\nu$, and die with rate $\mu>\nu$. The clone size distribution at steady state can be calculate and falls off rapidly (right).
{\bf B.} A minimal clonal selection model. Instead of dividing randomly, cells of the same clone all proliferate $m$-fold upon immune stimulation, which occurs with rate $r$. The clone size distribution of this process behaves as a power law for large clones. The exponent of the power law can be expressed as a function of the model parameters. {\bf C.-D.} The total number of clones can be expressed as a function of the model parameters for {\bf C)} the neutral model {\bf D)} the clonal selection model, and the total number of cells in the body, $T=3\cdot 10^{11}$. In the neutral model, the typical size of clones increases and diverges when division and death balance each other, $\mu\sim\nu$, leading to reduced diversity for a fixed number of cells. In the selection model, a similar divergence is observed as the power law exponent $\alpha$ gets close to 1.
}
\label{fig:numberclones}
\end{center}
\end{figure*}
}

\newcommand{\figurethree}{
\begin{figure*}
\begin{center}
\includegraphics[width=\linewidth]{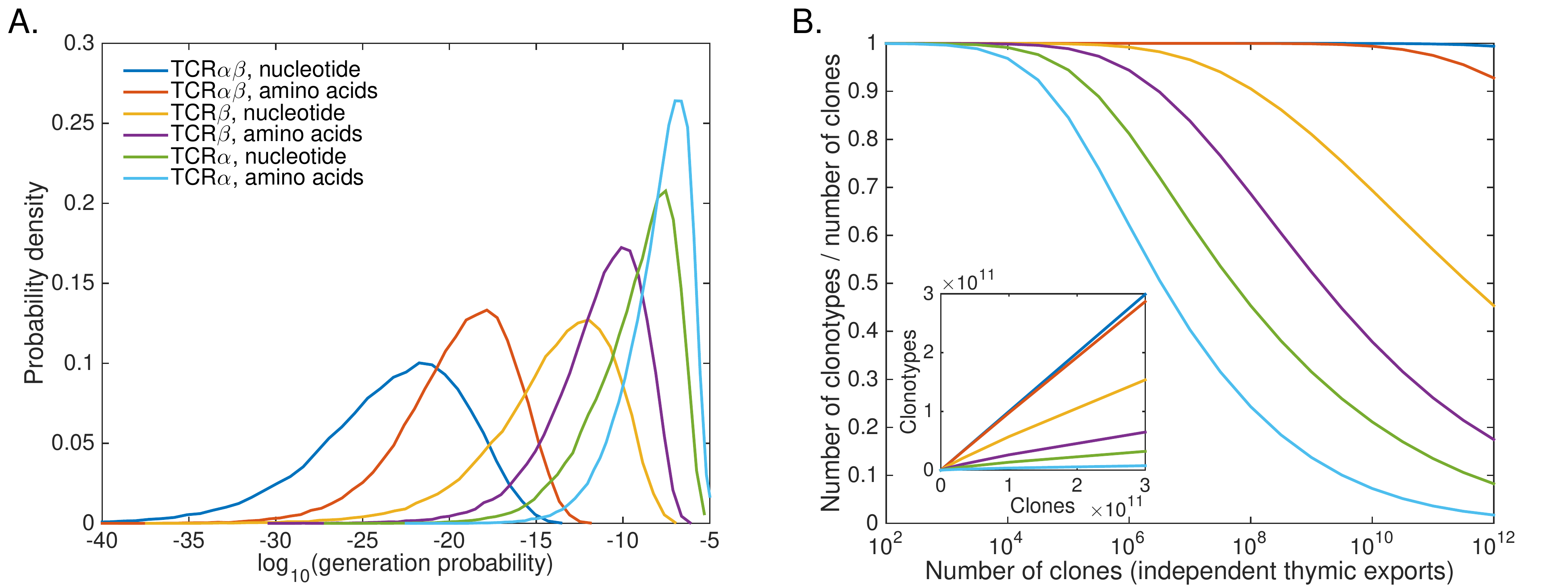}
\caption{\textbf{Convergent recombination. A.} Distributions of generation probabilities for the $\alpha$ and $\beta$ chains as well as $\alpha\beta$ pairs of T-cell receptors, for both nucleotide and amino-acid sequences, calculated using the OLGA software \cite{Sethna2019}. Most clonotypes have very low probability and are therefore unlikely to occur in two clones independently. High-probability clonotypes, however, will be generated several times in distinct T-cell clones (convergent recombination), reducing their diversity.  {\bf B.} Impact of convergent recombination on clonotype diversity. The ratio of the number of clonotypes to the number of clones is calculated using a model of recombination using OLGA, with the additional assumption that a random fraction $q$ of recombination events fail to pass selection \cite{Elhanati2018a}. This ratio
decreases as the number of clones increases, as redundant recombination events become more likely. The magnitude of this effect depends on the definition of clonotype (single chains or $\alpha\beta$ pairs, amino acids or nucleotides). It is small for full $\alpha\beta$ pairs. Inset: number of distinct clonotypes as a function of the number of clones. Selection parameter $q$: $q_{\alpha}=0.046$, $q_{\beta}=0.0091$, and $q_{\alpha\beta}=q_{\alpha}q_{\beta}$ (taken from Refs.~\cite{Elhanati2018a,Dupic2019a}).
}
\label{fig:clonotypesvsclones}
\end{center}
\end{figure*}
}

\begin{document}
\title{How many different clonotypes do immune repertoires contain?}

\author{Thierry Mora}
\affiliation{Laboratoire de physique de l'\'Ecole normale sup\'erieure
  (PSL University), CNRS, Sorbonne Universit\'e, and Universit\'e de
  Paris, 75005 Paris, France}
\author{Aleksandra M. Walczak}
\affiliation{Laboratoire de physique de l'\'Ecole normale sup\'erieure (PSL University), CNRS, Sorbonne Universit\'e, and Universit\'e de Paris, 75005 Paris, France}

\begin{abstract}
Immune repertoires rely on diversity of T-cell and B-cell receptors to protect us against foreign threats. The ability to recognize a wide variety of pathogens is linked to the number of different clonotypes expressed by an individual. Out of the estimated $\sim 10^{12}$ different B and T cells in humans, how many of them express distinct receptors? We review current and past estimates for these numbers. We point out a fundamental limitation of current methods, which ignore the tail of small clones in the distribution of clone sizes. We show that this tail strongly affects the total number of clones, but it is impractical to access experimentally. We propose that combining statistical models with mechanistic models of lymphocyte clonal dynamics offers possible new strategies for estimating the number of clones.
\end{abstract}

\maketitle

\section{Introduction}

The diversity of immune repertoires plays an important role in the host's ability to recognize and control a wide range of pathogens. While actual recognition of an antigen depends on having a relatively specific T-cell or B-cell receptor (TCR or BCR), multiple experimental examples show that reduced receptor diversity may limit the efficacy of adaptive immune repertoires \cite{Yager2008,Muraro2014}.
This effect becomes especially pronounced in individuals infected with cytomegalovirus \cite{Khan2002, Souquette2017, Smithery2018}, or with  immunosenescence~\cite{Qi2014}, when naive clonotypes are significantly reduced, and the organism is left to rely on reduced immune diversity.

The number of different clonotypes, called species richness in ecology, is thus an important quantity to estimate, both biologically and clinically. 
Here we review some of the experimental and theoretical approaches that have been used to estimate the number of distinct clonotypes in TCR and BCR repertoires, in naive, memory, or unfractionned repertoires.
Based on existing repertoire data and computational models, we demonstrate that no statistical method can overcome the limitations of small sampling. We argue that this problem, which is inherent to all existing methods, could be overcome by combining repertoire data with stochastic models of lymphocyte population dynamics, taking into account the caveats of convergent recombination and experimental noise.

\section{Past estimates}
The variable part of each TCR $\beta$ (and BCR heavy) chain is composed by putting together variable (V), diverse (D) and joining (J) regions. TCR $\alpha$ (and BCR light) chains only have V and J genes and no D genes.  
Additionally each chain experiences additions and deletions of nucleotides at the gene junctions, which increases the diversity. This junctional rearrangements have been identified as the main contributor to sequence diversity~\cite{Davis1988, Murugan2012}. The $\alpha$ and $\beta$ chain generations in TCR are separated in time, and have been shown to be independent~\cite{Dupic2019a,Grigaityte2017}. The total number of different $\alpha\beta$ pairs that the generation machinery can produce is {\em much} greater than the total number of receptors in the whole human population~\cite{Mora2016,Elhanati2018a,Dupic2019a}. Thus, each person harbors only a small fraction of the potential diversity of receptors. To estimate the number of distinct receptors, we need to count them.

An early quantitative direct estimate of the size of the TCR repertoire dates back to Arstila et al~\cite{Arstila1999}, following earlier considerations~\cite{Davis1988, Davis1998}. 
Their approach was to focus on a subclass of receptors (either $\alpha$ or $\beta$) with a specific 
V-J class and length, and sequence them using low-throughput methods. The number of different sequences in that subclass is then extrapolated back to the get the full diversity by dividing by the known frequency of V-J and length usage. The authors were very careful to verify their estimate in different V-J classes and donors, and to account for rare clonotypes that might have escaped sequencing.
A total of $\sim 10^{6}$ different TCR $\beta$ chains were thus estimated in a sample of $10^{8}$ T-cells.
Each $\beta$ chain paired with 25 different $\alpha$ chains, resulting in $\sim 25 \cdot 10^{6}$ distinct TCR$\alpha\beta$. Much smaller TCR $\alpha\beta$ diversity was reported in the memory subset, $\sim 2 \cdot 10^{5}$, consistent with the idea that memory cells form a selected and thus restricted subset.

With the onset of high-throughput sequencing of immune receptor repertoires~\cite{Weinstein2009,Robins2009,Boyd2009a,Six2013,Robins2013a,Georgiou2014,Heather2017,Minervina2019a,Bradley2019} came the realization of the importance of the sampling problem.
In sequenced repertoires, many clonotypes are seen just once, suggesting that there are possibly many more that have similar or slightly smaller sizes but were not sequenced, simply by chance. This issue does not only affect ``small'' clones. A clone of $10^{5}$ cells among a total of $3\cdot 10^{11}$ T cells will often not be seen even once in a typical sample of $10^6$ cells. To deal with this issue, a commonly adopted approach has been to use statistical estimators (see \cite{Laydon2015} for a overview).

Using the Poisson abundance statistical method \cite{Fisher1943}, Robins et al~\cite{Robins2009} obtained estimates of $\sim 10^{6}$ TCR$\beta$ nucleotide clonotypes for CD8 and CD4 naive cells, and $\sim 5 \cdot 10^{5}$-$10^6$ for CD8 and CD4 memory TCR$\beta$. 
Qi et al.~\cite{Qi2014} used another method called the Chao2 estimator~\cite{Chao2002}, which uses multiple replicates of the sequencing experiment, to estimate TCR species richness. 
They obtained much larger estimates than Robins et al.: $\sim 2 \cdot 10^8$ TCR$\beta$ nucleotide clonotypes in CD4 and CD8 naive repertoires, $\sim 1.5 \cdot 10^6$ clonotypes in CD4 memory repertoires, and about $5$-$10$ times fewer in CD8 memory. All those numbers decreased with age. 
Using the Poisson abundance method, diversities of $\sim 1$-$2\cdot 10^{9}$ for naive and  $\sim 5 \cdot10^{7}$-$10^{8}$ for memory BCR heavy chains were reported~\cite{DeWitt2016}.
Recent estimates of BCR heavy-chain species richness using an advanced statistical estimator \cite{Kaplinsky2016} yielded smaller diversities, ranging from $\sim 10^7$~\cite{Soto2019} to $10^7$-$10^9$~\cite{Briney2019}, presumably because they focused on amino acid rather than nucleotide clonotypes and ignored hypermutations in the V and J segments.

\figureone

\section{The sampling problem}
The approaches described above share the common problem that it is impossible to extrapolate what happens for small clones from small samples, which capture the largest clones \cite{Laydon2015,Kaplinsky2016}. Getting information on the small clonse is in fact impractical: it would require sequencing essentially all lymphocytes in an organism. Humans harbor of the order of $3\cdot 10^{11}$ T cells (and roughly the same order of B cells). Of these, only a few percents are contained in blood, of which a small fraction ($\sim 10^6$) is sampled in typical experiments (Fig.~\ref{fig:intro}A).
Even in mice, which contain fewer lymphocytes ($\sim 10^8$ T cells) and can be sacrificed to isolate all the body's lymphocytes, cell loss during the experiment hampers this approach.

In Fig.~\ref{fig:intro}B-D, we illustrate with simulations what happens when one analyses samples of $10^6$ cells from three synthetic repertoires. These repertoires are described by different clone size distributions, corresponding to a widely different number of clones: a pure power law (Fig.~\ref{fig:intro}B), a mixture of a power law and neutral model \cite{Hubbell2001} (Fig.~\ref{fig:intro}C), and a power law with a low-frequency cut-off (Fig.~\ref{fig:intro}D). Their species richness are widely different, ranging
from $N\sim 7\cdot 10^5$ to $1.6\cdot 10^{10}$. 
Yet, the sampled repertoires show similar clone size distributions, and comparable observed diversity ($10^5$-$10^6$), because they behave similarly for large clone sizes, but drastically differ in the tail of small clone sizes.

Any statistical method that extrapolates from observations assumes, knowingly or implicitly, an underlying model for how the clone size distribution behaves for the smallest clones. The Poisson abundance and Chao estimators \cite{Fisher1943,Chao1984,Chao2002} discussed earlier both assume a well peaked distribution of clone sizes, which is not the case in our examples. As a result, the Chao1 \cite{Chao1984} estimator can underestimate species richness by up to a 1,000-fold factor (Fig.~\ref{fig:intro}B). {\rev A more advanced estimator such as Recon \cite{Kaplinsky2016} gives even worst underestimates.}

Real repertoires are likely affected by this problem.
The sampled (large clone) part of their clone size distribution has been shown to follow a power law both for TCR and BCR \cite{Mora2016}. Naive subsets display shorter tails of large clones \cite{Oakes2017}, suggesting that the power-law behaviour in unfractioned repertoires is dominated by memory clones. Our three synthetic examples are consistent with power laws for large clones, but differ greatly for small clones, yielding very different species richness. Extrapolating the distribution of clone sizes is the key idea behind DivE \cite{Laydon2015a} and Recon \cite{Kaplinsky2016}, which were proposed to estimate diversity in TCR subsets. However, these approaches assume that the behaviour at large clones is informative for small clones, which may not always be true.

\figuretwo

\section{Proposed solution: stochastic modeling}

To access small clones that cannot be directly probed experimentally, we need to explicitly model the biological processes that shape these distributions, without having to take a leap of faith.
Unlike extrapolation, such models might predict behaviours for small clones that are quantitatively different than the trend suggested by large clones.
Of course, model assumptions should be tested experimentally, their parameters estimated from measurements, and confidence intervals put on their predictions.
We now briefly review two simple models that have been proposed to describe the dynamics and clone size distributions of naive and memory repertoires.

Cells in naive repertoires have not experienced strong proliferation due to antigen recognition. Nevertheless not all clones are of the same size, in part because clones leave the thymus with different initial sizes, and in part because they undergo stochastic division and death. The simplest model of naive repertoires is Hubbell's neutral model of ecology \cite{Hubbell2001}, which assumes constant division and death rates ($\nu<\mu$) for each cell, with new clones introduced with rate $\theta$ and constant initial size $k$ (Fig.~\ref{fig:numberclones}A, left). 
More complex variants of that model may include intrinsic fitness differences between clones or cells, e.g. through competition for self-antigens \cite{Lythe2016} or cytokines \cite{Desponds2016}.

Under that simple model, the steady state distribution of clones can be computed analytically \cite{Desponds2016,Greef2019,Altan-Bonnet2019}, and falls off exponentially for clones larger than $k$ (Fig.~\ref{fig:numberclones}A, right), meaning that large clones are rare. The total number of clones $N$ can also be calculated analytically as a function of the model parameters $\mu$, $\nu$, and $k$, as well as the total number of cells $T$ (Fig~\ref{fig:numberclones}C). Unless cell division almost exactly balances death ($\nu\sim\mu$), or the introduction clone size $k$ is large, the typical naive clone size is fairly small. This means that the total number of clones is very large, and comparable to the total number of cells. To get a more precise estimate would require to measure the division rate of naive T cells $\nu$, and initial clone size $k$.

A limitation of this approach is the assumption that the clone size distribution quickly reaches a steady state.
Naive T cells are very long-lived $\mu^{-1}\sim 3\ {\rm  years}$ \cite{DeBoer2013}, and the size of the naive pool changes with age, so that steady state may never be reached. Transient models of naive repertoires remain to be explored in more detail. A recent experimental study suggests that some T-cell naive clones are much larger than predicted by the neutral theory \cite{Greef2019}. However the origin of these outliers is not yet well understood, and may have to do with the inadequacy of our current definitions of naive and memory cells through surface markers.

Modeling memory repertoires requires taking into account the expansion and then contraction dynamics after an infection. These dynamics are driven by new pathogens that infect the host, are recognized and then cleared~\cite{DeBoer1995}, which leads to a constantly changing antigenic landscape~\cite{Desponds2016}. 
This random encounter with antigens can be simply modeled by bursts of division events for all cells of the same clone, with rate $r$, causing each cell to effectively multiply $m$ times into memory following antigen clearance (Fig.~\ref{fig:numberclones}B, left). Again, this model can be solved exactly in the continuous limit at steady state. The clone size distribution follows a power law for large clones (Fig.~\ref{fig:numberclones}B, right). The predicted number of clones $N$ depends critically on the power law exponent $\alpha$, and can be calculated as a function of the model parameters (Fig.~\ref{fig:numberclones}D). $N$ drops to zero for power law exponent $\alpha$ close to 1. Interestingly, measured exponents $\alpha$ from unfractioned T cell $\beta$ chain repertoires (whose large-clone tail is believed to be dominated by memory clones) range from $1$ to $\sim 1.5$ \cite{Mora2016}. This high sensitivity to parameters makes estimates from data very difficult.

Extensions of this model include the emergence of antigenic ``niches'' \cite{Desponds2017}, where clonal expansion is limited by antigen availability, leading to diminishing returns upon multiple stimulation events \cite{Mayer2019}. Such mechanisms would limit the size of the largest clones and would cut off the power law behaviour, which is not observed in data.

\figurethree

\section{Connecting models to data}
Several caveats and corrections must be taken into account when linking stochastic models of population dynamics such as discussed above to repertoire data. Of importances are the issues of convergent recombination and experimental noise.

Population dynamics models focus on clones, defined as the set of cells originating from a common recombination event. However, two recombination events can lead to exactly the same sequence. The two corresponding clones would be indistinguishable, and form a single clonotype in the repertoire.
This effect can be corrected for by using models of recombination \cite{Marcou2018,Sethna2019}. These models, which are inferred from data, can predict the distribution of generation probabilities of full receptors, or of single chains, both at the level of nucletoide or amino acid sequences, as shown in Fig.~\ref{fig:clonotypesvsclones}A for TCR. From this distribution, the probability of convergent recombination can be computed to predict the number of distinct clonotypes as a function of the number of ``clones'', defined as independent recombination events (Eq.~7 of \cite{Elhanati2018a}). In Fig.~\ref{fig:clonotypesvsclones}B we plot that prediction for the $\alpha$, $\beta$, and $\alpha\beta$ TCR in humans, for both nucleotide and amino acid clonotypes.
These computations show that, for the full TCR$\alpha\beta$ clonotypes, convergent recombination is so rare that it hardly affects species richness. For the $\alpha$ and $\beta$ chains alone, however, the effect is substantial. Since most repertoire data are of single chains, this correction should be applied when linking data to the type of models discussed above, as was done in Ref.~\cite{Greef2019}.

An additionnal issue complicates the comparison of models to data: experimental noise in the observed frequencies of clonotypes. 
In practice the number of reads (or unique molecular identifiers when they are used) $n$ observed in data for a given clonotype is not simply the result of random sampling, and is not
distributed according to a Poisson law, as was assumed in Fig.~\ref{fig:intro} and in all previous work on diversity estimation. Instead, noise is over-dispersed, due to additional noise caused by DNA amplification and library preparation prior to sequencing.
This noise model can be fitted using replicates of the repertoire sequencing experiment. This inference is impossible to separate from the inference of the clone size distribution $\rho(f)$. The two must thus be learned simultaneously from replicates by maximizing the likelihood of observed abundances, which depend on both the clone size distribution and the noise properties~\cite{Pogorelyy2018c}.
Applying this approach to the $\beta$ chain of unpartioned T cells with $\rho(f)\propto f^{-1-\alpha}$, yields species richness of $N\sim 10^8-10^9$, with power-law exponent $\alpha\approx 1$-$1.2$~\cite{Pogorelyy2018c}. In this estimate, the power law is taken as a given, and not linked to a model of clonal dynamics. A full mechanistic model treatment combined with the statistical model remains an interesting direction to explore, which could help shed light on the differences between memory and naive repertoires.

\section{Sampling and repertoire sharing}
Several recent papers have focused on shared immune receptors from high throughput BCR repertoire data \cite{Greiff2017a, Briney2019, Soto2019}. These high profile analyses report absolute percentages of shared clonotypes. However, it was shown in the context of TCR (but the same holds for BCR) that these fractions are not absolute properties of the repertoires, but rather depend on sampling depth and the number of individuals that share the clonotypes \cite{Mora2016,Elhanati2018a}. Sharing estimates based on samples of the repertoire are bound to grossly underestimate the true sharing fraction. Therefore, reporting sharing percentages without appropriate information about sample sizes is meaningless.

To assess the true overlap between the repertoires of two or more individuals, we would need to sequence all their lymphocytes, which is impractical.  However, statistical model of sequence probabilities can be used to extrapolate sharing estimates to the full repertoire size $N$, provided that number is known \cite{Elhanati2018a}. For instance, clonotypes whose probability $p$ is larger than $1/N$ are expected to be present in $1-e^{-pN}> 63\%$ of individuals, and can be considered ``public''. Recombination models such as the one of Fig.~\ref{fig:clonotypesvsclones}A can be used to estimate the fraction of clonotypes that are public. For example, for $N=10^{10}$, the model predicts that about $15\%$ of TCR$\beta$ amino acid clonotypes expressed by human individuals are public \cite{Elhanati2018a}.

\section{Conclusions}

While we have focused our review on the number of distinct clonotypes, what really matters for biological function is the number of different specificities. Due to cross-reactivity, each receptor can recognize many antigens and each antigen can be recognized by many receptors with different strengths. To account for this degeneracy, we would need to define a functional coverage of the antigenic space \cite{Zarnitsyna2013}. However, we currently do not have a comprehensive sequence-to-function maps for TCR and BCR that would allow us to estimate such a quantity.

Simply counting clonotypes also ignores their relative abundances. Clonotypes expressed by very few cells may not be as relevant for immune protection as very frequent clonotypes. Other diversity measures such as Hill numbers account for differences in frequencies \cite{Yaari2015,Mora2016}. Some of these measures are in fact more robust than species richness, because they put more focus on large clones and are less susceptible to sampling noise. Depending on the question, these  measures may be better suited than species richness.

Our discussion has focused mostly on T cells, and has ignored the complications of hypermutations in BCR, which cause lineages to split into many clonotypes. Whether diversity is defined at the level of lineages or clonotypes will lead to different answers \cite{DeWitt2016,Soto2019,Briney2019}. Developping specialized population dynamics models of B cell development and affinity maturation that include hypermutations is an interesting research direction.

We emphasized that estimating species richness cannot be disentangled from estimating the full distribution of clone sizes. As we gain insight into various aspects of lymphocyte dynamics, from thymic output to infection and memory formation, better mathematical descriptions can be leveraged to propose refined forms for the clone size distribution, and to fit their parameters to observations.
Only with such a combination of modeling and data will we be able to get a better picture of repertoire diversity and immune coverage.

{\bf Acknowledgements. 
} This work was partly supported by ERC CoG 724208.
\medskip

\bibliographystyle{pnas}

\begin{thebibliography}{10}

\bibitem{Yager2008}
Yager EJ, {et~al.}
\newblock (2008) {Age-associated decline in T cell repertoire diversity leads
  to holes in the repertoire and impaired immunity to influenza virus}.
\newblock \emph{J. Exp. Med.} 205:711--723.

\bibitem{Muraro2014}
Muraro P, Robins H
\newblock (2014) {T cell repertoire following autologous stem cell
  transplantation for multiple sclerosis}.
\newblock \emph{J. Clin. Invest.} 124:1168--1172.

\bibitem{Khan2002}
Khan N, {et~al.}
\newblock (2002) {Cytomegalovirus seropositivity drives the CD8 T cell
  repertoire toward greater clonality in healthy elderly individuals.}
\newblock \emph{J. Immunol.} 169:1984--1992.

\bibitem{Souquette2017}
Souquette A, Frere J, Smithey M, Sauce D, Thomas PG
\newblock (2017) {A constant companion: immune recognition and response to
  cytomegalovirus with aging and implications for immune fitness}.
\newblock \emph{GeroScience} 39:293--303.

\bibitem{Smithery2018}
Smithey MJ, Venturi V, Davenport MP, Buntzman AS, Vincent BG
\newblock (2018) {Lifelong CMV infection improves immune defense in old mice by
  broadening the mobilized TCR repertoire against third-party infection}.
\newblock \emph{PNAS} 115.

\bibitem{Qi2014}
Qi Q, {et~al.}
\newblock (2014) {Diversity and clonal selection in the human T-cell
  repertoire.}
\newblock \emph{Proc. Natl. Acad. Sci. U. S. A.} 111:13139--44.

\bibitem{Davis1988}
Davis MM, Bjorkman PJ
\newblock (1988) {T-cell antigen receptor genes and T-cell recognition.}
\newblock \emph{Nature} 334:395--402.

\bibitem{Murugan2012}
Murugan A, Mora T, Walczak AM, Callan CG
\newblock (2012) {Statistical inference of the generation probability of T-cell
  receptors from sequence repertoires}.
\newblock \emph{Proc. Natl. Acad. Sci.} 109:16161--16166.

\bibitem{Dupic2019a}
Dupic T, Marcou Q, Walczak AM, Mora T
\newblock (2019) {Genesis of the $\alpha$$\beta$ T-cell receptor}.
\newblock \emph{PLoS Comput. Biol.} 15:e1006874.

\bibitem{Grigaityte2017}
Grigaityte K, {et~al.}
\newblock (2017) {Single-cell sequencing reveals $\alpha$$\beta$ chain pairing
  shapes the T cell repertoire}.
\newblock \emph{bioRxiv:213462}.

\bibitem{Mora2016}
Mora T, Walczak AM
\newblock (2018) in \emph{Systems Immunology: An Introduction to Modeling
  Methods for Scientists}, eds{} Das J, Jayaprakash C
\newblock (CRC Press).

\bibitem{Elhanati2018a}
Elhanati Y, Sethna Z, Callan CG, Mora T, Walczak AM
\newblock (2018) {Predicting the spectrum of TCR repertoire sharing with a
  data-driven model of recombination}.
\newblock \emph{Immunol Rev} 284:167--179.

\bibitem{Arstila1999}
Arstila TP, {et~al.}
\newblock (1999) {A direct estimate of the human alphabeta T cell receptor
  diversity.}
\newblock \emph{Science} 286:958--961.

\bibitem{Davis1998}
Davis MM, {et~al.}
\newblock (1998) {Ligand recognition by alpha beta T cell receptors.}
\newblock \emph{Annu. Rev. Immunol.} 16:523--44.

\bibitem{Weinstein2009}
Weinstein JA, Jiang N, White RA, Fisher DS, Quake SR
\newblock (2009) {High-throughput sequencing of the zebrafish antibody
  repertoire.}
\newblock \emph{Science (80-. ).} 324:807--810.

\bibitem{Robins2009}
Robins HS, {et~al.}
\newblock (2009) {Comprehensive assessment of T-cell receptor beta-chain
  diversity in alphabeta T cells.}
\newblock \emph{Blood} 114:4099--4107.

\bibitem{Boyd2009a}
Boyd SD, {et~al.}
\newblock (2009) {Measurement and clinical monitoring of human lymphocyte
  clonality by massively parallel {\{}VDJ{\}} pyrosequencing}.
\newblock \emph{Sci Transl Med} 1:12ra23.

\bibitem{Six2013}
Six A, {et~al.}
\newblock (2013) {The past, present and future of immune repertoire biology -
  the rise of next-generation repertoire analysis}.
\newblock \emph{Front. Immunol.} 4:413.

\bibitem{Robins2013a}
Robins H
\newblock (2013) {Immunosequencing: applications of immune repertoire deep
  sequencing}.
\newblock \emph{Curr. Opin. Immunol.} 25:646--652.

\bibitem{Georgiou2014}
Georgiou G, {et~al.}
\newblock (2014) {The promise and challenge of high-throughput sequencing of
  the antibody repertoire.}
\newblock \emph{Nat. Biotechnol.} 32:158--68.

\bibitem{Heather2017}
Heather JM, Ismail M, Oakes T, Chain B
\newblock (2017) {High-throughput sequencing of the T-cell receptor repertoire:
  pitfalls and opportunities}.
\newblock \emph{Brief. Bioinform.} 19:554--565.

\bibitem{Minervina2019a}
Minervina A, Pogorelyy M, Mamedov I
\newblock (2019) {TCR and BCR repertoire profiling in adaptive immunity}.
\newblock \emph{Transpl. Int.} pp 0--2.

\bibitem{Bradley2019}
Bradley P, Thomas PG
\newblock (2019) {Using T Cell Receptor Repertoires to Understand the
  Principles of Adaptive Immune Recognition}.
\newblock \emph{Annu. Rev. Immunol.} 37:547--570.

\bibitem{Laydon2015}
Laydon DJ, Bangham CRM, Asquith B, Crm B
\newblock (2015) {Estimating T-cell repertoire diversity: limitations of
  classical estimators and a new approach}.
\newblock \emph{Philos Trans R Soc Lond, B, Biol Sci} 370:20140291.

\bibitem{Fisher1943}
Fisher R, {Steven Corbet} A, Williams C
\newblock (2016) {The Relation Between the Number of Species and the Number of
  Individuals in a Random Sample of an Animal Population}.
\newblock \emph{J. Anim. Ecol.} 12:42--58.

\bibitem{Chao2002}
Chao A, Bunge J
\newblock (2002) {Estimating the number of species in a stochastic abundance
  model.}
\newblock \emph{Biometrics} 58:531--539.

\bibitem{DeWitt2016}
DeWitt WS, {et~al.}
\newblock (2016) {A Public Database of Memory and Naive B-Cell Receptor
  Sequences}.
\newblock \emph{PLoS One} 11:e0160853.

\bibitem{Kaplinsky2016}
Kaplinsky J, Arnaout R
\newblock (2016) {Robust estimates of overall immune-repertoire diversity from
  high-throughput measurements on samples}.
\newblock \emph{Nat. Commun.} 7:1--10.

\bibitem{Soto2019}
Soto C, {et~al.}
\newblock (2019) {High frequency of shared clonotypes in human B cell receptor
  repertoires}.
\newblock \emph{Nature} 566:398--402.

\bibitem{Briney2019}
Briney B, Inderbitzin A, Joyce C, Burton DR
\newblock (2019) {Commonality despite exceptional diversity in the baseline
  human antibody repertoire}.
\newblock \emph{Nature} 566:393--397.

\bibitem{Hubbell2001}
Hubbell SP
\newblock (2001) \emph{{The Unified Neutral Theory of Biodiversity and
  Biogeography}}.

\bibitem{Chao1984}
Chao A
\newblock (1984) {Nonparametric estimation of the number of classes in a
  population}.
\newblock \emph{Scand J Stat.} 11:265--270.

\bibitem{Oakes2017}
Oakes T, {et~al.}
\newblock (2017) {Quantitative Characterization of the t Cell Receptor
  Repertoire of Na{\"{i}}ve and Memory subsets Using an Integrated experimental
  and Computational Pipeline Which Is Robust, Economical, and Versatile}.
\newblock \emph{Frontiers in genetics} 8:1--17.

\bibitem{Laydon2015a}
Laydon DJ, {et~al.}
\newblock (2014) {Quantification of HTLV-1 Clonality and TCR Diversity}.
\newblock \emph{PLoS Computational Biology} 10:1--13.

\bibitem{Lythe2016}
Lythe G, Callard RE, Hoare RL, Molina-par{\'{i}}s C
\newblock (2016) {How many TCR clonotypes does a body maintain ?}
\newblock \emph{Journal of Theoretical Biology} 389:214--224.

\bibitem{Desponds2016}
Desponds J, Mora T, Aleksandra W
\newblock (2016) {Fluctuating fitness shapes the clone size distribution of
  immune repertoires}.
\newblock \emph{Proc Natl Acad Sci USA} 113:274.

\bibitem{Greef2019}
Greef PCD, {et~al.}
\newblock (2019) {The naive T-cell receptor repertoire has an extremely broad
  distribution of clone sizes}.
\newblock \emph{bioRxiv:691501}.

\bibitem{Altan-Bonnet2019}
Altan-Bonnet G, Mora T, Walczak AM
\newblock (2019) {Quantitative Immunology for Physicists}.
\newblock \emph{arXiv:1907.03891}.

\bibitem{DeBoer2013}
{De Boer} RJ, Perelson AS
\newblock (2013) {Quantifying T lymphocyte turnover}.
\newblock \emph{J. Theor. Biol.} 327:45--87.

\bibitem{DeBoer1995}
{De Boer} RJ, Perelson aS
\newblock (1995) {Towards a general function describing T cell proliferation.}
\newblock \emph{J. Theor. Biol.} 175:567--576.

\bibitem{Desponds2017}
Desponds J, Mayer A, Mora T, Walczak AM
\newblock (2017) {Population dynamics of immune repertoires}.
\newblock pp 1--9.

\bibitem{Mayer2019}
Mayer A, Balasubramanian V, Walczak AM, Mora T
\newblock (2019) {How a well-adapting immune system remembers}.
\newblock \emph{PNAS} 116:8815--8823.

\bibitem{Marcou2018}
Marcou Q, Mora T, Walczak AM
\newblock (2018) {High-throughput immune repertoire analysis with IGoR}.
\newblock \emph{Nat. Commun.} 9:561.

\bibitem{Sethna2019}
Sethna Z, Elhanati Y, Callan CG, Walczak AM, Mora T
\newblock (2019) {OLGA: fast computation of generation probabilities of B- and
  T-cell receptor amino acid sequences and motifs}.
\newblock \emph{Bioinformatics} btz035.

\bibitem{Pogorelyy2018c}
Pogorelyy MV, {et~al.}
\newblock (2018) {Precise tracking of vaccine-responding T-cell clones reveals
  convergent and personalized response in identical twins}.
\newblock \emph{Proc Natl Acad Sci} 115:12704--12709.

\bibitem{Greiff2017a}
Greiff V, {et~al.}
\newblock (2017) {Systems Analysis Reveals High Genetic and Antigen-Driven
  Predetermination of Antibody Repertoires throughout B Cell Development}.
\newblock \emph{Cell Rep.} 19:1467--1478.

\bibitem{Zarnitsyna2013}
Zarnitsyna VI, Evavold BD, Schoettle LN, Blattman JN, Antia R
\newblock (2013) {Estimating the diversity, completeness, and cross-reactivity
  of the T cell repertoire}.
\newblock \emph{Front. Immunol.} 4:485.

\bibitem{Yaari2015}
Yaari G, Kleinstein SH
\newblock (2015) {Practical guidelines for B-cell receptor repertoire
  sequencing analysis}.
\newblock \emph{Genome Med.} 7:121.

\end{thebibliography}

\end{document}